\documentclass[journal]{IEEEtran}
\ifCLASSINFOpdf
\else
\fi
\usepackage{graphicx}
\usepackage{float}
\usepackage{amssymb}
\usepackage{amsmath}
\usepackage{cite}
\usepackage{multirow}
\usepackage{stfloats}
\usepackage{algorithm, algorithmic}

\begin{document}
%\linenumbers
\title{Self-supervised SAR-optical Data Fusion and Land-cover Mapping using Sentinel-1/-2 Images}
\author{Yuxing~Chen,~\IEEEmembership{~}
	    Lorenzo~Bruzzone,~\IEEEmembership{Fellow,~IEEE}% <-this % stops a space
        
\thanks{Y. Chen and L. Bruzzone are with the Department of Information Engineering and Computer Science, University of Trento, 38122 Trento, Italy (e-mail:yuxing.chen@unitn.it;lorenzo.bruzzone@unitn.it).}
\thanks{Corresponding author: L. Bruzzone}}

\maketitle

%\begin{spacing}{1.5}
\begin{abstract}
%o
The effective combination of the complementary information provided by the huge amount of unlabeled multi-sensor data (e.g., Synthetic Aperture Radar (SAR) and optical images) is a critical topic in remote sensing.
%O
Recently, contrastive learning methods have reached remarkable success in obtaining meaningful feature representations from multi-view data.
%O
However, these methods only focus on image-level features, which may not satisfy the requirement for dense prediction tasks such as land-cover mapping.
%O
In this work, we propose a self-supervised framework for SAR-optical data fusion and land-cover mapping tasks.
%O
SAR and optical images are fused by using multi-view contrastive loss at image-level and super-pixel level in the early, intermediate and later fashion individually.
%O
For the land-cover mapping task, we assign each pixel a land-cover class by the joint use of pre-trained features and spectral information of the image itself.
%o
Experimental results show that the proposed approach achieves a comparable accuracy and that reduces the dimension of features with respect to the image-level contrastive learning method.
%O
Among three fusion fashions, the intermediate fusion strategy achieves the best performance.
%O
The combination of the pixel-level fusion approach and spectral indices leads to further improvements on the land-cover mapping task with respect to the image-level fusion approach, especially with few pseudo labels.
\end{abstract}

%\IEEEpeerreviewmaketitle

\section{Introduction}
%O
%\begin{spacing}{1.5}
\IEEEPARstart{E}{very} year a large number of Earth Observation Satellites are operated to monitor human activities, Earth environment, and their mutual influences across our planet.
%O
Hundreds of terabytes of remote sensing (RS) data are accumulated per day from various systems, which cover most bands of the electromagnetic spectrum and include both active and passive sensors \cite{chi2016big}.
%O
In this context, deep learning methods, especially supervised deep learning approaches, have been developed to process and analyze such massive amounts of multimodal RS data for specific applications, such as land-cover mapping, target recognition, and change detection.
%O
However, these applications are mostly limited to the use of a single type of image and require a large amount of labeled data for the training of the algorithm.

%O
The most common are deep learning techniques based on single modality data, e.g., multispectral, hyperspectral, LiDAR, and Synthetic Aperture Radar (SAR).
The fusion of various RS data from different sensors has not received sufficient attention yet.
%O
However, it is well known that the complementary use of multimodal RS data offers more complete information on a scene that can result in better performance in many applications \cite{gomez2015multimodal}. 
%o
For example, multispectral/hyperspectral images acquire spectral information provided to interpret the land-cover categories on the basis of their spectral signatures, while radar images provide dielectric properties, and are not affected by cloud occlusions.

%o
Inspired by the success of deep learning in computer vision (CV), some works \cite{audebert2018beyond,jiang2020change,bermudez2018sar,li2018hyperspectral,srivastava2019understanding,ienco2019combining,bermudez2018sar} have investigated the fusion of multimodal RS data using deep learning methods.
Their results have shown that deep learning techniques play a significant role in multimodal RS data fusion.
%o
However, recent successes of deep learning techniques in multimodal RS data fusion are mainly focused on supervised methods, which are often limited from the availability of annotated data. 
%o
Labeled remote sensing data are often scarce.
The limited access to such labeled data has driven the development of unsupervised methods, such as generative models (e.g., GAN\cite{goodfellow2014generative}, CAE\cite{masci2011stacked}, VAE\cite{kingma2013auto}), which are probably the most used representation learning methods.
%o
Nevertheless, some studies have shown that such generative models overly focus on pixels rather than on abstract feature representations.

%O
For land-cover mapping tasks, existing supervised methods \cite{gargiulo2020integration,liu2019integration,ienco2019combining} often require prohibitive annotation effort for labeling data.
%C
Thus, semi-supervised techniques \cite{wu2020semi} have been developed and explored that require very little labeled data to train the classifier.
%C
Self-training is a popular semi-supervised technique that trains the supervised classifier in various iterations with the help of a few labels and the large pool of unlabeled samples \cite{aydav2019self}.
%O
Besides, Chen et al \cite{ChenKSNH20}, find that the use of self-supervised pretraining for fine-tuning in self-training can give a nice improvement in computer vision.% and this is very common in NLP.
%O
Unlike semi-supervised learning, many unsupervised land-cover mapping works \cite{aswatha2020unsupervised,li2019novel} have used a set of spectral indices (e.g., normalized difference water index (NDWI), normalized difference vegetation index (NDVI), bare soil index (BI) and backscattering values (BS)) of SAR and optical images to select training samples from the image itself.
%C
These indices have the potential to highlight a set of land-cover classes in terms of their physical characteristics.
%O
We state that can be used to select pseudo labels in a self-straining paradigm.

%o
Recent researches \cite{bachman2019learning,oord2018representation,tian2019contrastive,hjelm2018learning,chen2016infogan} in contrastive learning demonstrate how these methodologies can encourage the network to learn more interpretable and meaningful feature representations.
%o
This resulted in improvements in classification and segmentation tasks, where contrastive methods outperformed the generative counterparts.
%o
However, existing contrastive methods (MCL) rarely consider the pixel-wise feature representations.

%o
To address this limitation, in this paper we propose a new self-supervised fusion approach to fuse the complementary information present in SAR and optical images at the pixel level.
%O
The proposed approach can be implemented according to three fusion strategies: I) early fusion (PixEF), II) intermediate fusion (PixIF) and late fusion (PixLF).
%O
We also investigate the self-trained land-cover classification in considering spectral information in SAR-optical images and the benefits of self-supervised pre-trained features.
%linear protyco??
The proposed SAR-optical fusion approach is compared with the instance level contrastive method under the common linear protocol in the context of a land-cover mapping task.
%O
The evaluation of the self-supervised land-cover mapping approach was performed on a subset of DFC2020.

%o 
The main contributions of this work are as follows:
%o

1) We first introduce and verify the effectiveness of multi-view contrastive loss in SAR-optical data fusion.
Then we propose a self-supervised approach, which can obtain pixels-wise feature representations from SAR and optical image pairs without using any annotation.
%o
This is achieved by using U-Net and contrastive loss, preserving local information at the superpixel level.
%o 

2) We investigate the impact of different fusion strategies (i.e., early fusion, intermediate fusion and later fusion) in the proposed approach.
Concretely, late and intermediate fusion fashion learns feature representations by comparing SAR and optical images directly, while early fusion fashion distills the complementary information from the concatenation of image pairs. 
In addition, the efficiency of SAR-optical fusion with respect to the use of a single modality in the land-cover mapping task is analyzed correspondingly.

%o
3) We conduct a self-supervised land-cover mapping task by jointly using the spectral information of SAR-optical images and the pretraining fusion features.

%o
The rest of this paper is organized as follows.  
%o
Section II presents the related works of SAR-optical data fusion based on supervised and unsupervised methods.   
%o
Section III introduces the proposed approach by describing the three considered self-supervised data fusion fashion and a self-training land-cover mapping task. 
%o
The descriptions of the dataset, network setup, and experimental settings are given in Section IV.
Experimental results obtained on linear protocol and a land-cover mapping task are illustrated in Section V. 
%o
Finally, Section VI concludes the paper.

\section{Related Works}

%o
Various feature fusion methods, including supervised learning and unsupervised learning techniques, have been investigated to improve the performance of combining complementary information from SAR and optical images.
%o
Early works in both already proved the effectiveness of combining SAR and optical data with the multi-layer perception (MLP) classifier \cite{bruzzone1997multisource,bruzzone1999neural}.
%o
Newly Sentinel-1 and Sentinel-2 images are combined to improve land-use and land-cover (LULC) classification accuracy on monsoon regions using a random forest model in \cite{steinhausen2018combining}.
%o
However, these fusion algorithms are a simple concatenation of SAR and optical images and have no capability to learn high-level features. 
%o
With the development of deep convolution neural networks (CNNs), hand-crafting features have become unnecessary for SAR-optical feature fusion.
%o
Task-relevant features can be extracted from the input data automatically during the network training.
%o
Kussul \textit{et al.}\cite{kussul2017deep} first explore the deep CNNs in SAR-optical fusion for LULC classification and demonstrate its superior to traditional MLP classifiers.
%o
In \cite{feng2019integrating}, Feng \textit{et al.} propose a multi-branch CNNs to improve the classification accuracy in coastal areas by fusing Sentinel-1 and Sentinel-2 images.
%O
A multi-temporal W-Net was proposed to integrate Sentinel-1 and Sentinel-2 images in land-cover mapping \cite{gargiulo2020integration}.
%O
An integration of fully convolutional networks (FCN) and object-based methods are proposed for LULC mapping and achieved an overall accuracy of 95.33\% with the Sentinel SAR and optical images \cite{liu2019integration}.
%o
Recently, Dino \textit{et al.} \cite{ienco2019combining} propose a deep learning architecture, namely TWINNS, to fuse Sentinel-1 and Sentinel-2 time series data in land-cover mapping.
%O
Besides the complete supervised fashion, Wang \textit{et al.} \cite{wang2020weakly} find a correlation between the image classification accuracy and the segmentation accuracy, and demonstrate that a model trained on image classification performs well on a segmentation task in a weakly supervised way.

%o
Supervised learning methods, despite their powerful ability to feature learning of SAR-optical images, are often limited by the amount of annotated data.
%o
An alternative method is to use unsupervised learning techniques, such as principal component analysis (PCA), convolutional AutoEncoder (CAE) and canonical correlation analysis (CCA)\cite{hotelling1992relations}.
These techniques can learn feature representations from unlabeled multimodal data.
%o
In \cite{amarsaikhan2010fusing}, Amarsaikhan \textit{et al.} use PCA to enhance the features extracted from SAR-optical images and improve the urban land-cover maps.
%o
Similarly, multi-view learning methods also provide a solution to the unsupervised combination of complementary information from SAR-optical images.
In \cite{geng2017classification}, Jie \textit{et al.} propose a deep bimodal autoencoder (BDAE) to fuse SAR and multispectral images for classification. 
In \cite{nielsen2017canonical}, Allan \textit{et al.} jointly analyze Sentinel SAR and optical data for change detection using CCA.
%o
Based on CCA, Andrew \textit{et al.} propose the deep canonical correlation analysis (DCCA), which learns separate representations for each modality from a shared latent subspace using CNNs.

\begin{figure*}[pt]
	\centering
	\includegraphics[width=7.0in]{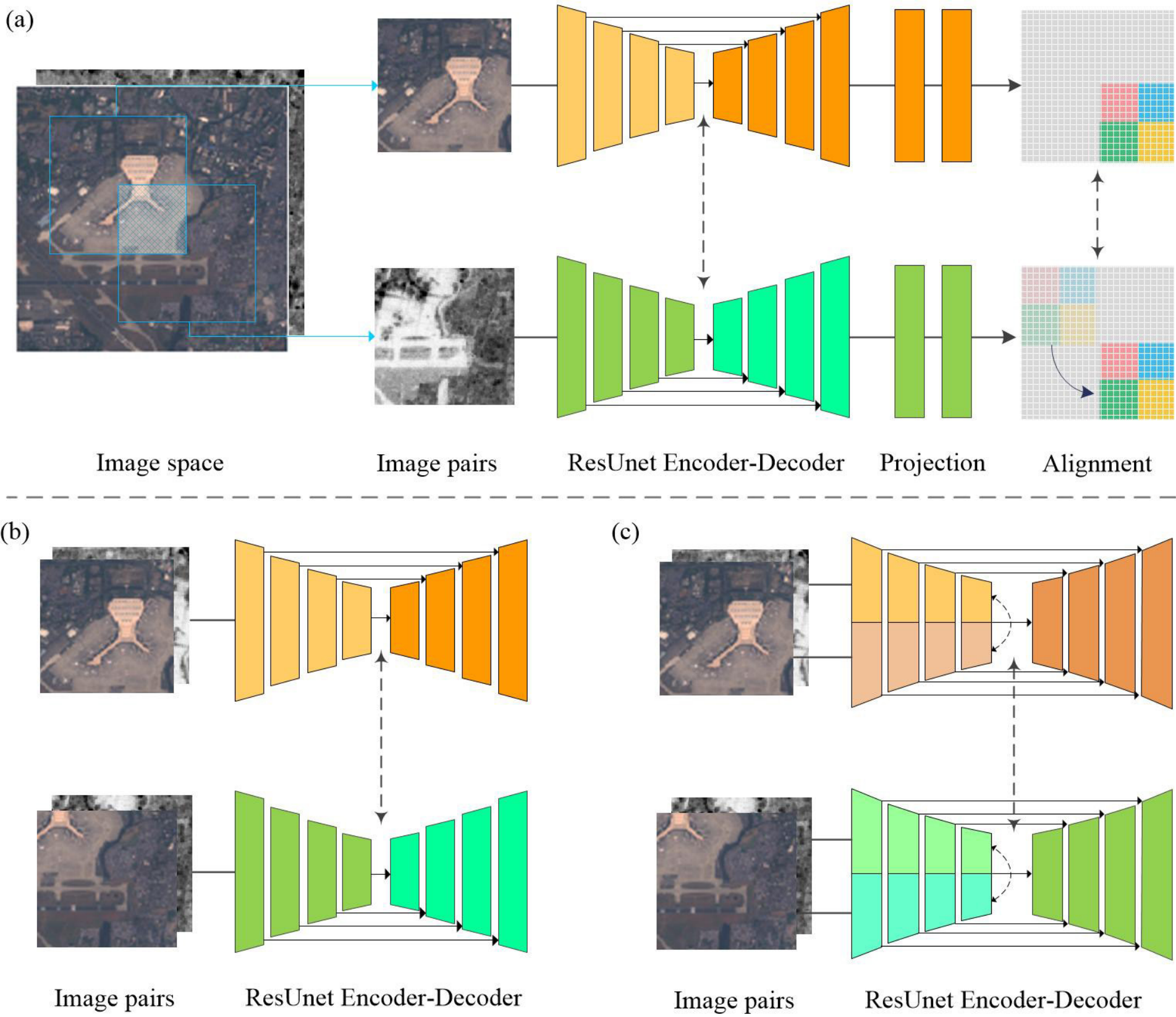}
	\caption{Overview of the presented self-supervised SAR-optical fusion approach. The dash arrow line represents a contrastive loss. (a) An illustration of pixel-wise representation learning framework for the late fusion fashion. The two inputs have an offset but keep an overlap. The approach follows the common contrastive learning architecture where both branches consist of a ResUnet block and a projection. Then, a shift transformation is included in the one branch for aligning representations between two branches. (b) The ResUnet block follows the early fusion fashion. (c) The ResUnet block follows the intermediate fusion where the encoder contains two parts used for encoding SAR and optical images independently.}
	\label{fig1}
\end{figure*}

%o
Despite the success of these methods, their performance is limited in feature learning ability or overly focus on the pixel values.
%O
The recent success of contrastive learning \cite{oord2018representation} shows the possibility of alleviating such a dependency and obtaining useful representations from unlabeled data.
%O
Tian \textit{et al.} \cite{tian2019contrastive} propose an extension of the contrastive loss to self-supervised learning at the multi-view setting, where different views are related to the same context.
%o
Maximizing mutual information between different modalities within the same context will force the corresponding sub-networks to generate the same or similar features.

%O
Even with the useful representation, the absence of labels is not enough for the land-cover mapping task.
%o
Many works \cite{li2019novel,aswatha2020unsupervised,wu2020semi} have been dedicated to decreasing the use of labels using the spectral information in SAR-optical images themselves.
%O
Aswatha \textit{et al.} \cite{aswatha2020unsupervised} jointly use scattering and spectral properties from SAR and multi-spectral images to distinguish land-cover classes as training samples.
Besides, they train a random forest classifier to predict land-cover classes based on the selected samples and learned features. 
%o
Wu \textit{et al.}\cite{wu2020semi} utilize self-training to gradually assign highly confident pseudo labels to unlabeled hyperspectral images by clustering and employs spatial constraints to regulate the self-training process.
%o
Rather than training from scratch, Chen \textit{et al.} find that the self-training based on self-supervised pre-training can improve the performance in classification task \cite{chen2020big}.

\section{Methodology}
%o
This section presents the proposed self-supervised approach to SAR-optical image fusion, which aims to learn pixel-wise representations from unlabeled SAR-optical image pairs.
%O
The proposed SAR-optical fusion approach consists of three fusion strategies: early fusion, intermediate fusion and late fusion (see Fig. \ref{fig1} (a, b and c)).
%O
Afterward, a land-cover mapping task based on the spectral information of the image itself and the pre-trained features is presented.

\subsection{Network Architecture}
%O
The proposed approach has two branches (Fig. 1 (a)), where the input image of each branch has a relative shift.
%O
Each branch contains a ResUnet block followed by a linear layer projector.
%O
After the projection, the same shift operation is performed on the output for feature alignment between two branches.
%o
We adopt a similar ResUnet architecture as the \cite{zhang2018road} and only used residual blocks in the encoder part.
%C
Like U-net, ResUnet consists of encoder, decoder and skip connections between downsampling and upsampling path.
%O
In this work, ResNet-18 is used as the encoder of the ResUnet block, but without the fourth residual block.
%O
The decoder part has three blocks and consists of a convolution layer (Conv) and batch normalization (BN), ReLU, and upsampling in each block.
%O
All padding type in the ResUnet block was changed as the same padding.
%O
The projector consists of a $1\times1$ Conv for each pixel and follows with the ResUnet block is used to reconstruct the learned representations.

%O
Feature alignment is achieved by using same shift operation on input images and output features of two branches individually.
%O
Specifically, given an input $I_1$ consisting of SAR and optical image pairs obtained from the same scene, we use a shift operation to make a random offset of $I_1$ along the x and y-axis direction.
%O
In this way, we can define the random shift transformation as $T$ and obtain an augmented view $I_2=T(I_1)$.
%O
During the training, the augmented view and original input are fed into two branches, respectively, to obtain pixel-wise representations $v_1$ and $v_2$.
%O
To align pixel-wise representations of two branches, the same transformation is applied to the output of the other branch $v_1=T(v_1)$.

%o
In particular, two branches of PixEF and PixIF share the same parameters, but the encoder of PixIF is split into two groups and each group has half channels of the counterpart of PixEF in each layer.
%O
We denote the network of PixEF as $U_e$ and its encoder as $E_e$.
Meanwhile, the network of PixIF as $U_i$ and its two indepent encoders as $E_{i1}$ and $E_{i2}$.
Unlike these two models, the PixLF has two independent branches with half channels of PixEF in each layer, where the input channels were adjusted to the input images.
We denote the network of PixLF as $U_l$ and its two indepent encoders as $E_{l1}$ and $E_{l2}$.

\subsection{Loss Function}
%O
The proposed approach consists of two types of contrastive loss based on image and superpixel individually.
%o
The main idea behind a contrastive loss is to find a feature representation that is invariant to augmentations.
%o
Given a dataset of $S$ that consists of a collection of image pairs $\{s_1^i,s_2^i\}_{i=1,2,\dots,N}$, we consider each image pair $(s_1^i,s_2^i)$ sampled from the joint distribution $p(s_1^i,s_2^i)$, which we call positives.
%o
And $s_2^j$ is taken from other scene, then samples $(s_1^i,s_2^j)$ from the product of marginals $p(s_1^i)p(s_2^j)$, which we call negatives.
%o
The model $h(\cdot)$ is expected to know which pair was drawn from the joint distribution while the other is not exactly by computing their cosine similarity with a hyper-parameter $\tau$.
%o
In multi-view setting, the model $h(\cdot)$ is a neural network consits of two braches that with the independent parameter $f_{{\theta}_1}$ and $f_{{\theta}_2}$.
\begin{equation}\label{eq1}
h\left(s_{1}, s_{2}\right)=\exp \left(\frac{f_{\theta_{1}}\left(s_{1}\right) \cdot f_{\theta_{2}}\left(s_{2}\right)}{\left\|f_{\theta_{1}}\left(s_{1}\right)\right\| \cdot\left\|f_{\theta_{2}}\left(s_{2}\right)\right\|} \cdot \frac{1}{\tau}\right)
\end{equation}
%o
The final loss function can be writen as $L(f_{{\theta}_1}, f_{{\theta}_2}, S)$ given dataset $S$.
\begin{equation}\label{eq2}
L(f_{{\theta}_1}, f_{{\theta}_2}, S)=-\underset{S}{\mathbb{E}}{\left[\log \frac{h(s_1^1, s_2^1)}{\sum_{j=1}^{N} h(s_1^1, s_2^j)}\right]}
\end{equation}
where $(s_1^1,s_2^1)$ is a positive pair sample, $(s_1^1,s_2^j|j\ge 1)$ is a negative pair sample and $S=\{s_1^1, s_2^1,s_2^2, \cdots, s_2^{N-1}\}$ is a set that contains $N-1$ negative samples and one positive sample.
%o
In the training process, the network is trained to increase the value of positive pairs and decrease the value of negative pairs.
This results in a feature representation that is close to positive pairs whereas it is not for negative pairs.

%o
In the pixel-level contrastive loss, we sample and average features from two branches over superpixels that are located on the overlap between two branches.
%O
This aims to keep the consistency of the normalized pixel-wise representations between two branches.
%O
Here, we construct a set of pixel-wise feature pairs $P$ where positive feature pair $(p_1^i,p_2^i)$ is sampled from the same location while $p_2^j$ in negative pairs was taken from another location.
%O
Compared with the instance-level contrastive learning, this loss function can make the model get more detailed representations and is more suitable for dense prediction downstream tasks.  
%O spixel
To overcome the noise when using single pixels, we adopt the contrastive loss at the superpixel level.
%O
Together with the pixel-level contrastive loss, an instance-level contrastive loss is used to improve the performance.
%O
The instance-level loss help to discriminate the similarity between the shited views.
%O
Like pixel-level loss, we can construct a set $M$ of concatenated image pairs, where $(m_1^i,m_2^i)$ is sampled from the same scene $i$ while $m_2^j$ is taken from another scene.
%O
Finally, we use the pixel-wise contrastive loss in conjunction with the image-level contrastive loss, leading to the total loss of three fusion approches:
\begin{equation}
\begin{array}{l}
\mathcal{L}_{\text {e}}=L(U_e, U_e, P) + L(E_e,E_e, M) \\
\mathcal{L}_{\text {l}}=L(U_{l1},U_{l2},P) + L(E_{l1}, E_{l2}, M) \\
\mathcal{L}_{\text {i}}=L(U_i, U_i, P) + L(E_{i1}, E_{i2}, M) + L(E_{i}, E_{i}, M)
\end{array}
\end{equation}
where $\mathcal{L}_{\text {e}}$,$\mathcal{L}_{\text {l}}$ and $\mathcal{L}_{\text {i}}$ are the loss function of PixLF, PixEF and PixIF individually.

\subsection{Self-training Land-cover mapping}
%O
To further implement the proposed SAR-optical fusion approach in land-cover mapping tasks that do not rely on manually annotated data, a self-training process is introduced and used in conjunction with the spectral information of images.  
%O
We first calculate a set of spectral indices of each image and then select reliable, representative and diverse pseudo labels according to the characteristics of different land-cover classes.
%O
For different spectral indices, the larger or minor index value the pixel has, the greater possibility that pixels are associated with the related land-cover classes.
%O
In this paper, NDVI, NDWI, BI and BS were used to produce the index images for different classes.
%O
The generation of pseudo labels for different land-cover classes that not only decided on the certain spectral index but also on those indices related to other classes, which can be utilized for synergistically labeling a pixel.

%O
To avoid choosing a threshold value for each index, we use k-means \cite{krishna1999genetic} to over cluster Sentinel-1/-2 images and calculate the average spectral indices and average backscatter value for each class.
Here, we segment sentinel-2 images to eight classes and sentinel-1 images to four classes.
%O
Then, we define a set of generic rules (see Algorithm 1) for labeling 6 classes considering all indices and the combination of pseudo-classes defined by k-means on Sentinel-1 and -2 images.
We denote labeled pixels of the forest, grassland, water, urban, bare land, and sparse vegetation as $U^f$, $U^g$, $U^w$, $U^u$, $U^b$ and $U^s$.
%O
For instance, the NDVI can indicate forest and grassland, and backscatter values in SAR images can further discriminate between these two classes. 
%O
It is noted that just a few pixels with high confidence (thresholds) are labeled with a given class and the mean value of pre-defined classes is used as the threshold value.

\begin{figure}[pt]
	\centering
	\includegraphics[width=3.5 in]{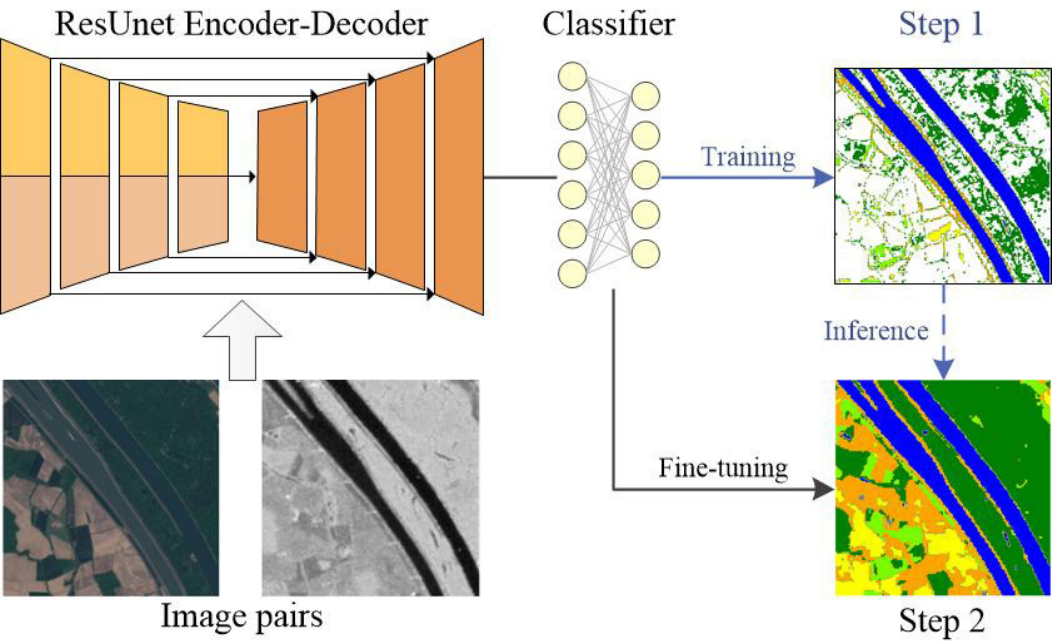}
	\caption{Overview of the presented self-training land-cover mapping approach. The step one is training on the sparse pseudo labels first. Then, training whole network on the predicted labels in step two.}
	\label{fig2}
\end{figure}

%O
The generated sparse pseudo labels are utilized to train a linear classifier layer based on the learned fusion features and then the classes of unlabeled pixels are predicted using the well-trained models.
%C
However, the generated labels inevitably contain incorrect predictions.
%O
To overcome this problem, we fine-tune all parameters on the inference labels.
%o
The training framework in this task is illustrated in Fig. \ref{fig2}.
%o
It is noted that the pre-trained PixIF model is used as the backbone and followed a linear classifier.

\begin{algorithm}
	\renewcommand{\algorithmicrequire}{\textbf{Input:}}
	\renewcommand{\algorithmicensure}{\textbf{Output:}}
	\renewcommand{\algorithmicelsif}{\textbf{elif}}
	\caption{Unsupervised Training Samples Collection.}
	\label{alg:1}
	\begin{algorithmic}[1]
		\REQUIRE $I_1$, $I_2$ images, NDVI, NDWI, BI and BS
		\ENSURE $U^{f}$, $U^{g}$, $U^{w}$, $U^{u}$, $U^{b}$, $U^{s}$
		\STATE Overclustering $I_1$, $I_2$ into $m$ and $n$ classes.
		\STATE Selecting the classes with maximun indicies values $H_{ndvi}$, $H_{ndwi}$, $H_{bi}$ in $I_2$ and $H_{bs}$ in $I_1$ as well as the class with minum value $L_{bs}$ in $I_1$. Also selecting the classes with medium NDVI and BI value$M_{ndvi}$ and $M_{bi}$. Then calculate their mean values $V_{ndvi}$, $V_{ndwi}$, $V_{bi}$, $V_{bs}$.
		\STATE Iterating each pixel $p$ in $I_1$ and $I_2$.
		\IF{$p$ in $H_{ndwi}$ and $L_{bs}$; NDWI$(p_2) > V_{ndwi}$} 
		\STATE $p \in U^{w}$
		\ELSIF {$p$ in $H_{ndvi}$ and $H_{bs}$; NDVI$(p_2) > V_{ndvi}$} 
		\STATE $p \in U^{f}$
		\ELSIF {$p$ in $H_{ndvi}$ and $L_{bs}$; NDVI$(p_2) > V_{ndvi}$} 
		\STATE $p \in U^{g}$
		\ELSIF {$p$ in $H_{bs}$; BS$(p_1) > V_{bs}$} 
		\STATE $p \in U^{u}$
		\ELSIF {$p$ in $H_{bi}$ and $L_{bs}$; BI$(p_2) > V_{bi}$} 
		\STATE $p \in U^{b}$
		\ELSIF {$p$ in $M_{bi}$ and $M_{ndvi}$; NDVI$(p_2) < V_{ndvi}$ } 
		\STATE $p \in U^{s}$
		\ELSE
		\STATE $p \in None$
		\ENDIF
		\STATE \textbf{return} $U^{f}$, $U^{g}$, $U^{w}$, $U^{u}$, $U^{b}$, $U^{s}$
	\end{algorithmic}  
\end{algorithm}

\section{Design of Experiments}
%O
In this section, we present the dataset for training and validation of the proposed approach as well as the land-cover mapping task.
%O
Besides, the details of network setup and evaluation experiments are introduced.

\subsection{Description of the Dataset}

\subsubsection{DFC2020}
We developed our experiments on the DFC2020 dataset.
%o
The DFC2020, which has been issued by the IEEE-GRSS 2020 Data Fusion Contest \cite{yokoya20202020}, is used as the training set and the evaluation set for comparison between different methods. 
%o
This dataset consists of a total of 6114 quadruple samples, which are SAR-optical image pairs, MODIS-derived labels, and more accurate semi-manually derived high resolution (10m) land-cover maps \cite{schmitt2020weakly}.
%o
SAR images were acquired by Sentinel-1 and consist of dual-polarized (VV and VH) components and the optical images were taken by multi-spectral Sentinel-2.
%o
Each pixel in the DFC2020 was assigned to a land-cover class manually, which has eight fine-grained classes (i.e., Forest, Shrub-land, Grassland, Wetlands, Croplands, Urban/built-up, Barren and Water).
%O
We also provide the image-level label for each scene, which is derived by the majority class of the related pixel-level land-cover maps.
%o
The previous research \cite{wang2020weakly} pointed out the effectiveness of training a CNN on image-level labels that can guide the weakly supervised model (WSL) to learn a powerful representation of images.

\subsubsection{Training and Test Splits}
%o
A random split of the DFC2020 dataset into a training set (1000) and a test set (5114) were applied in this work.
%O
Here, the training set is used to tune the parameters of the linear protocol in the evaluating phase.
%o
The test set was used to validate the effectiveness of the features learned using different methods.
%o
To assess the effectiveness of these methods with limited labels, we randomly split the training set into five groups with 10, 50, 100, 200, and 1000 samples.  
Each small number group is a sub-samples of the corresponding full training set. 
%o
Note that all self-supervised and unsupervised models were trained on unlabeled SAR-optical image pairs.
%O
In addition, we selected 2000 samples from the whole data set for the self-training land-cover mapping task.
%O
We reclassify the cropland, wetland, and grassland as grassland and reclassify shrubland and barren as bared land.
Other classes keep the same as the original.
%O
Therefore, this data set only contains 6 classes (i.e., Forest, Grassland, urban, bare land, water and sparse vegetation) according to the land surface properties.

\subsection{Network Setup}
%o
The training process of the self-supervised approach includes three parts for PixEF, PixLF and PixIF.
%O
For PixEF, SAR-optical image pairs were concatenated as one input.
For PixIF and PixLF, SAR and optical images are in input to two branches independently. 
%O
The Adam with a learning rate of 0.0003, a weight decay of 0.0004 and a momentum of 0.9 was adapted as our optimizer.
%C
We use a mini-batch size of 1000; models are run for 700 epochs.
%O
We deploy a step scheduling learning rate policy in the training process.
%o
Apart from the shift transformation, we further apply a random flip transformation to improve the performance of the proposed approach.

%o
For self-training in the land-cover mapping task, we train the model with a mini-batch size of 50, and models are run for 100 epochs.
%O
Adam optimizer is used with an initial learning rate of $1e^{-4}$ for the pre-trained encoder layer and $3e^{-4}$ for the other parameters.
%C
When finetuning, the initial learning rate was set as $1e^{-4}$ for all parameters.

\subsection{Experiment Settings}
%O
To evaluate the learned feature representation of different methods, we provide an evaluation with a linear classifier followed by the frozen features on the test set.
%O
In particular, the feature representation in the proposed approach has 256 channels, while that of the comparison methods (i.e., DCCA and MCL) is a concatenation of multi-level features with 512 channels.
%O
Note that we decided to use a linear classifier as our main evaluation metric for the quality of representations since it has a small number of extra parameters.
%O
The learning rate is set to 0.05 and the SGD with a mini-batch size of 8 was adopted as the optimizer for the linear protocol as well as the maximum number of epochs is 50.

%O
In this experiment, we also assess the performance of the proposed approach applied to the Sentinel-1 (S1) image, the Sentinel-2 (S2) image, and both of them (S1S2).
%O
This is done to validate whether the SAR-optical fusion can obtain more discriminative representations than single image type for the downstream land-cover mapping task.
%O
To this purpose, we train the proposed approach only on single modality images in the proposed early fusion fashion.
%O
Then the linear protocol on the frozen pre-trained models was used to evaluate the effectiveness of learned representations with limited training labels.

%O
In the land-cover mapping task, we adopt features obtained from the proposed self-supervised SAR-optical fusion approach and instance-level contrastive approach as pre-training.
%O
However, the instance level self-supervised method does not have an encoder and can not perform the same strategy.
%O
To provide a fair comparison, we also adopt the same decoder and the classifier of the proposed approach as a readout, which is followed by encoders pre-trained by instance self-supervised methods and used to reconstruct the concatenation features for downstream tasks.
%O
Furthermore, the proposed approach is fine-tuned with the predicted pseudo labels.
%O
To present a sparse label scenario, we just keep less than 10-pixel labels for each class of each image, where aband the class with less than 10-pixel labels.

\section{Results}

\subsection{Linear Evaluation}
\begin{figure}[pt]
	\centering
	\includegraphics[width=3.0in]{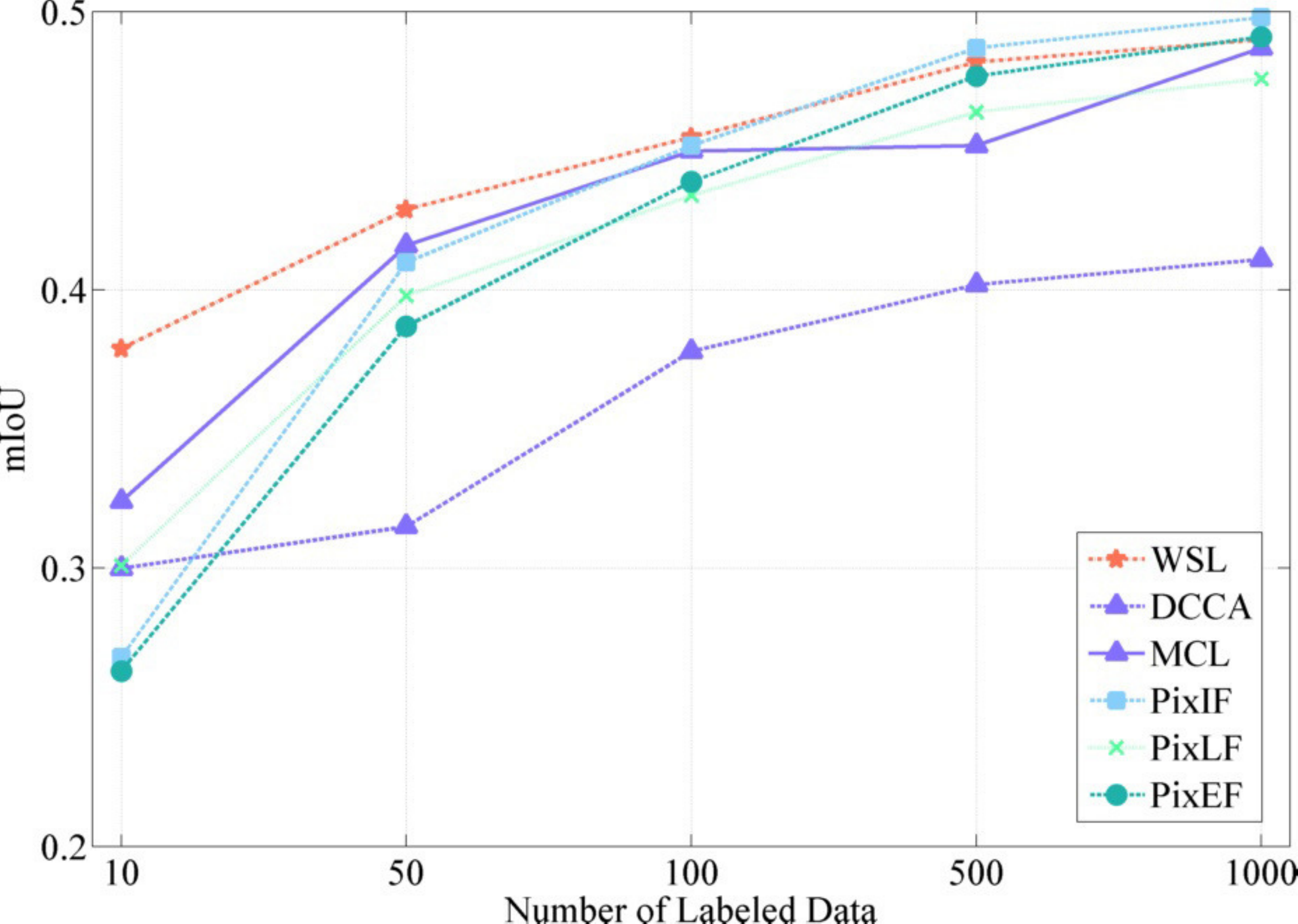}
	\caption{Mean intersection over union metric achieved by different methods on test set versus the number of samples used for the training of the linear classifier on frozen encoders.}
	\label{fig3}
\end{figure}

%o
The performance of the proposed approaches (PixEF, PixIF and PixLF) were evaluated on the test set in comparison to the weakly supervised method (WSL) and the other two instance-level self-supervised methods (DCCA and MCL).
%O
Also in this case we considered different amounts of labeled data for the training of the linear classifier (see Fig. \ref{fig3}).
%o
The average class accuracy (AA) and mean intersection over union (mIoU) are common metrics used to assess the performance in land-cover mapping and are used to evaluate the overall precision of all land-cover classes in our work.

%o
Fig. \ref{fig3} shows the linear protocol results on mIoU.
As we can see that the proposed PixIF outperforms all other methods while the DCCA performs significantly worse against any other method when the number of training samples increases.
%o
However, the WSL method outperforms all other methods with few labels.
%O
In general, the proposed PixEF and PixLF as well as the MCL have a similar performance.
And the gap between WSL and all contrastive approaches is reduced when the number of labeled samples increases.
%O
Finally, the proposed PixIF outperforms WSL in the case of 1000 training samples.
%O
Among contrastive approaches, the performance of PixIF is slightly better than MCL.
It is worth noting that MCL obtained representations with 512 channels, while that of proposed approaches only with 256 channels. 
This means that the proposed approach significantly reduces the dimensionality of features while keeping or even improving the feature representation ability.

%\end{spacing}

\begin{table}[pt]
	\centering
	\caption{Class-wise and overall accuracies achieved on the test set by a linear classifier used with the different methods considering 1000 SAR-optical training samples.}
	\label{table1}
	\renewcommand\tabcolsep{5.0pt}
	\centering
	\begin{tabular}{ccccccc}
		\hline
		\multirow{2}{*}{Class} & \multicolumn{6}{c}{Average Accuracies \%} \\ \cline{2-7}
		           &  WSL & DCCA & MCL  & PixIF& PixLF& PixEF \\ \hline
		Forest     & 90.2 & 92.7 & 90.9 & 92.6 & 92.3 & 91.3\\
		Shrub-land & 70.4 & 26.3 & 40.4 & 50.8 & 46.3 & 53.2\\
		Grassland  & 57.6 & 63.0 & 68.7 & 73.0 & 64.5 & 74.5\\
		Wetlands   & 70.4 & 56.9 & 64.9 & 62.7 & 57.2 & 59.9\\
		Croplands  & 81.2 & 72.8 & 82.8 & 84.7 & 85.8 & 80.0\\
		Urban      & 86.6 & 86.9 & 87.7 & 87.8 & 84.4 & 87.1\\
		Barren     & 34.5 & 6.3  & 46.3 & 30.9 & 29.5 & 37.0\\
		Water      & 99.2 & 99.0 & 99.3 & 99.3 & 99.4 & 99.2\\ \hline
		AA         & 73.8 & 63.0 & 72.7 & 72.7 & 69.9 & 72.8\\ 
		mIoU       & 0.490& 0.411& 0.487& 0.498& 0.476&0.490\\\hline
	\end{tabular}
\end{table}

\begin{figure*}[pt]
	\centering
	\includegraphics[width=6.5in]{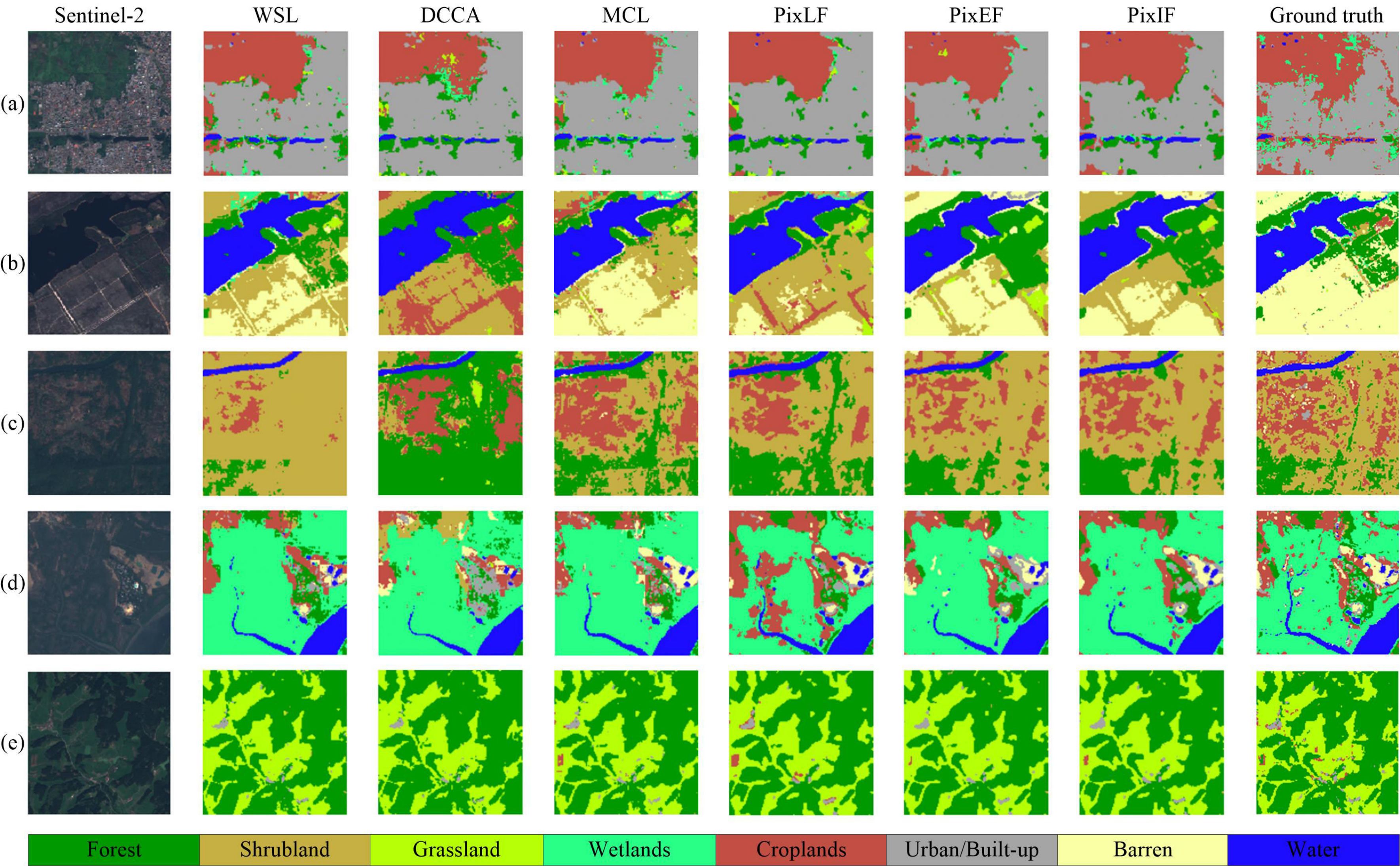}
	\caption{Land-cover maps achieved on five different images by different considered methods with a linear classifier. (see Table \ref{table1} for quantitative results.)}
	\label{fig4}
\end{figure*}

%o
Table \ref{table1} presents a detailed comparison of the class-by-class accuracy obtained by the classifier on the test set for each method when trained with 1000 labeled samples.
%o
According to the results, the proposed PixIF and PixEF and WSL achieve an AA higher than 72\% and the mIoU over 0.49 on the test set.
They sharply outperform DCCA, which obtain an AA smaller than 65\% and the mIoU less than 0.42. 
Among proposed approaches, PixIF obtained a higher mIoU while PixLF get the worst performers. 
%o
In addition to the quantitative evaluations, we also provide a qualitative visual comparison of the land-cover maps predicted by different methods.
%o
Fig. \ref{fig4} illustrates five examples of results. 
%o
Each example includes the predicted land-cover maps of different methods as well as the ground truth in DFC2020.
%o
As one can observe, the proposed PixIF, PixEF and WSL show better results than the rest of the methods in all cases, and the proposed PixIF confirms to be more effective than the other two fusion strategies (PixEF and PixLF).
%o
For each land cover class, similar conclusions to those derived by Table \ref{table1} can be given.

%O
In general, the results obtained in all comparisons confirm that the contrastive approach is superior to DCCA in this land-cover mapping task.
%o
The effectiveness of the proposed PixIF against other methods is due to its ability to use three-level contrastive loss.
%o
It is interesting to note that despite the annotations are used in WSL, the proposed approach achieves comparable performance without any use of labels.
%O
This confirms the effectiveness of the self-supervised methods in feature representation learning.

%O
We further investigate the performance of a single sensor image in the land-cover mapping task, where we train the proposed PixEF approach on Sentinel-1 (S1) and Sentinel-2 (S2) image as well as the concatenation of Sentinel-1/-2 (S1S2) images individually.
%o
Table\ref{table2} shows a quantitative evaluation of the accuracy on each class considered S1, S2 and S1S2 with the 1000 training samples under both two evaluations.
As one can see, the SAR-optical fusion outperforms the use of any single modality data in both linear protocol and fine-tune evaluations.
The performance of PixEF on S2 is very close to the performance on S1S2.
%o
While the performance on S1S2 has an increase of more than 10\% of AA with respect to the performance on S1 in both types of evaluation.
%o
In addition, the use of sentinel-2 data results in higher classification accuracy on all classes than the use of only sentinel-1 data for both evaluation methods.
%o。
For individual classes, water achieved the highest accuracy in all conditions, whereas the barren has the lowest accuracy.
%o
The classification accuracy of water does not show an obvious improvement after fusion, because there is already enough information in each single modality data.
However, the SAR-optical fusion improved the performance of shrubland, grassland and barren.

\begin{table}[pt]
	\centering
	\caption{Class-wise and overall accuracies achieved by PixEF on Sentinel-1 images alone (S1), Sentinel-2 images alone (S2) and Sentinel-1/-2 image fusion (S1S2) with the linear protocol and the fine-tuning evaluation.}
	\label{table2}
	\renewcommand\tabcolsep{8.0pt}
	\centering
	%0.625mm
	%\setlength{\tabcolsep}{0.625mm}{
	\begin{tabular}{ccccccc}
		\hline
		\multirow{2}{*}{class} & \multicolumn{3}{c}{Linear Evaluation} & \multicolumn{3}{c}{Fine-tuning Evaluation} \\ \cline{2-7} 
		& S1   & S2   & S1S2&  S1  & S2   & S1S2 \\ \hline
		Forest    & 84.2 & 90.3 & 91.3 & 86.2 & 92.2 & 92.0 \\
		Shrubland & 27.5 & 48.9 & 53.2 & 31.2 & 44.2 & 62.3 \\
		Grassland & 61.1 & 67.2 & 74.5 & 61.9 & 68.0 & 78.0 \\
		Wetlands  & 35.0 & 58.8 & 59.9 & 51.5 & 62.3 & 62.3 \\
		Croplands & 66.0 & 81.4 & 80.0 & 71.1 & 79.6 & 85.7 \\
		Urban     & 78.8 & 86.5 & 87.1 & 84.7 & 89.7 & 86.1 \\
		Barren    & 3.5  & 30.0 & 37.0 & 7.8  & 30.7 & 38.6 \\
		Water     & 98.8 & 99.2 & 99.2 & 99.2 & 99.5 & 99.4 \\ \hline
		Average   & 56.9 & 70.0 & 72.8 & 61.7 & 70.8 & 75.6 \\
		mIoU      & 0.362& 0.470& 0.490& 0.395& 0.474& 0.521\\ \hline
	\end{tabular}
	%}
\end{table}
\begin{figure}[pt]
	\centering
	\includegraphics[width=3.5in]{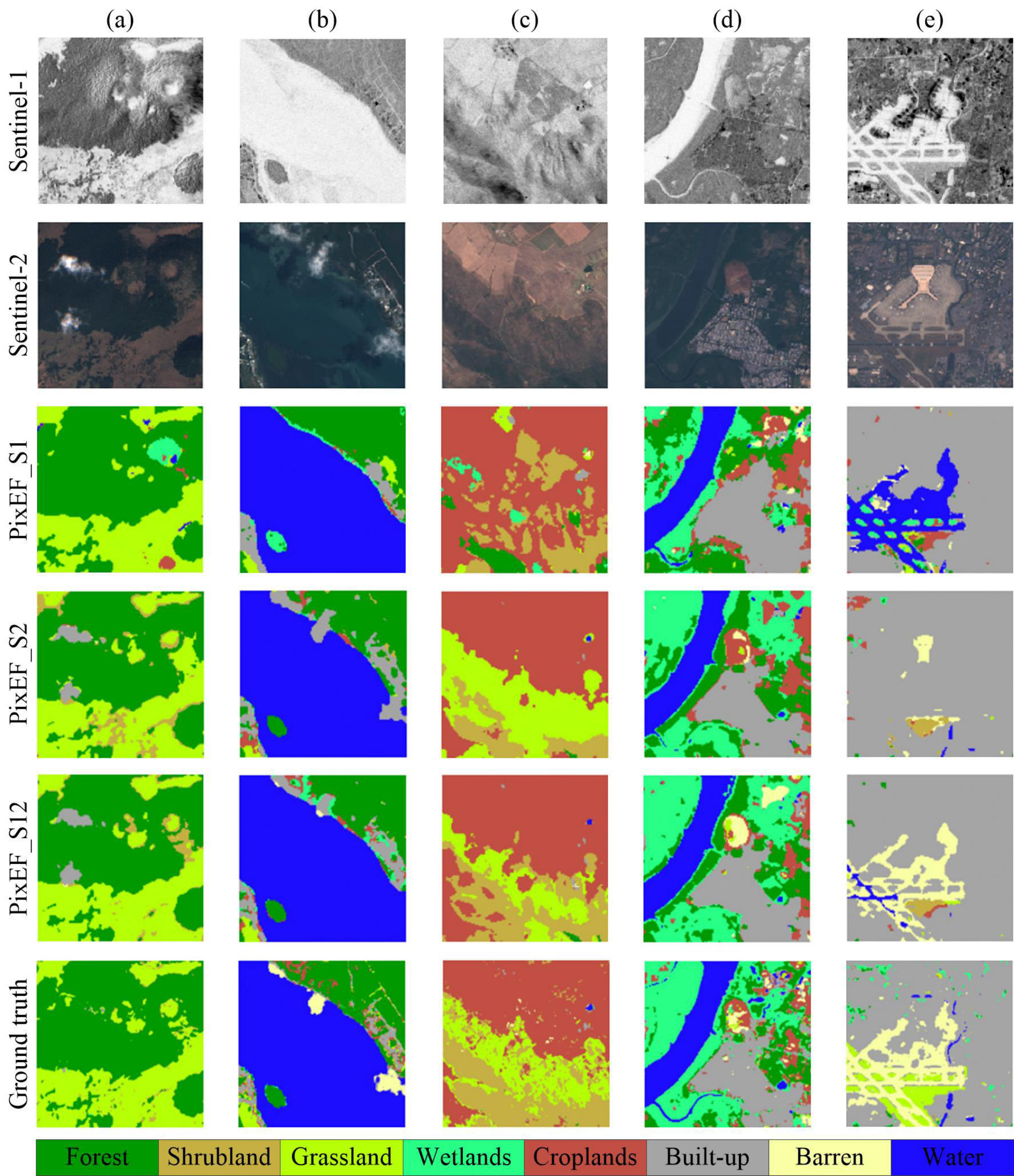}
	\caption{Land-cover maps obtained by PixEF on Sentinel-1 images alone (S1), Sentinel-2 images alone (S2) and  Sentinel-1/-2 image fusion with the linear classifier and fine-tuning evaluation for five different images. (see Table IV for quantitative results.)}
	\label{fig5}
\end{figure}

%o
Apart from quantitative assessment, we also made a visual quantitative comparison of both two evaluations on S1, S2 and S1S2, respectively. 
%o
Obviously, the performance of each modality and SAR-optical fusion keeps consistent in both evaluation criteria.
%o
As shown in Fig. \ref{fig5}, the SAR-optical fusion classifies various classes in a more accurate way, especially in barren, which obtains a significant improvement with respect to the use of single modality data.
%o
Besides, a trend can be figured out, that is, the methods with the input of fused S1S2 data achieve more smooth parsing results compared with the input of single modality data. 
%o
Moreover, Fig. \ref{fig5} also shows the advantages of each single modality and other improvements after SAR-optical fusion.
%o
Both Fig. \ref{fig5}(a) and Fig. \ref{fig5}(b) show that the presence of clouds on Sentinel-2 (but of course not in Sentinel-1) and the obvious influence on the corresponding classification map.
%o
The presence of the cloud leads to misclassification in Sentinel-2, but not in Sentinel-1.
%o
Although the misclassification induced by clouds is still present in the classification maps of SAR-optical fusion, it is significantly mitigated compared to the result of Sentinel-2 alone.
%o
A similar phenomenon is also present in Fig.\ref{fig5}(e), where the airport was distinguished as water in the Sentinel-1 result.
This is the results of the similar backscatter between flat runway and water.
%o
Conversely, this was correctly distinguished as built-up in the Sentinel-2 result.
%o
After SAR-optical fusion, the misclassification error in each modality was obviously reduced.
%o
Overall, the visual comparison is coherent with the quantitative results presented in Table \ref{table2} and confirms again the effectiveness of the presented self-supervised approach.

\subsection{Self-trained land-cover mapping task}
%o
\begin{table}[pb]
	\centering
	\caption{Class-wise and overall accuracies achieved on the test set by a linear classifier used with the different methods considering 2000 samples.}
	\label{table3}
	\renewcommand\tabcolsep{3.0pt} % 调整表格列间的宽度
	\centering
	\begin{tabular}{ccccccccc}
		\hline
		Method & Forest & Sparse Veg. & Grass. & Built. & Barren & Water & AA & mIoU\\ \hline
		MCL   & 87.3 & 36.9 & 53.3 & 68.8 & 42.5 & 99.0 & 64.7 & 0.408 \\ \hline
		PixIF & 89.6 & 39.3 & 55.7 & 73.0 & 47.4 & 99.2 & 67.4 & 0.438 \\ \hline
		Fine. & 91.6 & 38.2 & 57.2 & 74.4 & 47.1 & 99.3 & 68.0 & 0.449 \\ \hline
	\end{tabular}
\end{table}

We further validate the proposed approach on a particular self-supervised land-cover mapping task.
Table\ref{table3} presents a quantitative evaluation of the accuracy on each class considered image-level feature fusion (MCL) and the pixel-level feature fusion (PixIF) as well as the fine-tuning results of PixIF.
%O
As one can see, the pixel-level fusion approach achieved a higher AA of 67.4\% sharply outperforming the image-level fusion approach, which obtain an AA of 64.7\%. 
Similarly, compared with the image-level fusion, the pixel-level fusion outperforms it with a margin of 0.03 in mIoU.
%o
In particular, the fine-tuning results obtain a higher AA and mIoU, with a significant improvement of 3.3 \% and 0.04 with respect to the image-level fusion.

%o
For individual classes, forest and water achieved the highest accuracy.
%o
On the contrary, the accuracy of sparse vegetation and barren is below 50\%.
%o 
The classification of the sparsely vegetated ground is a big challenge for all methods.
%o
The possible reason is that the spectral characteristics of forest and water are very distinctive and very easy to be distinguished.
%O
On the contrary, barren and sparse vegetation present a high similarity of textural and spectral characteristics.
%o
Another possible reason has reported in \cite{yu2020spatial}, is that there are also many misclassifications among shrub-land, wetland, barren, and cropland contained in the origin ground-truth label.

\begin{figure}[pt]
	\centering
	\includegraphics[width=3.2in]{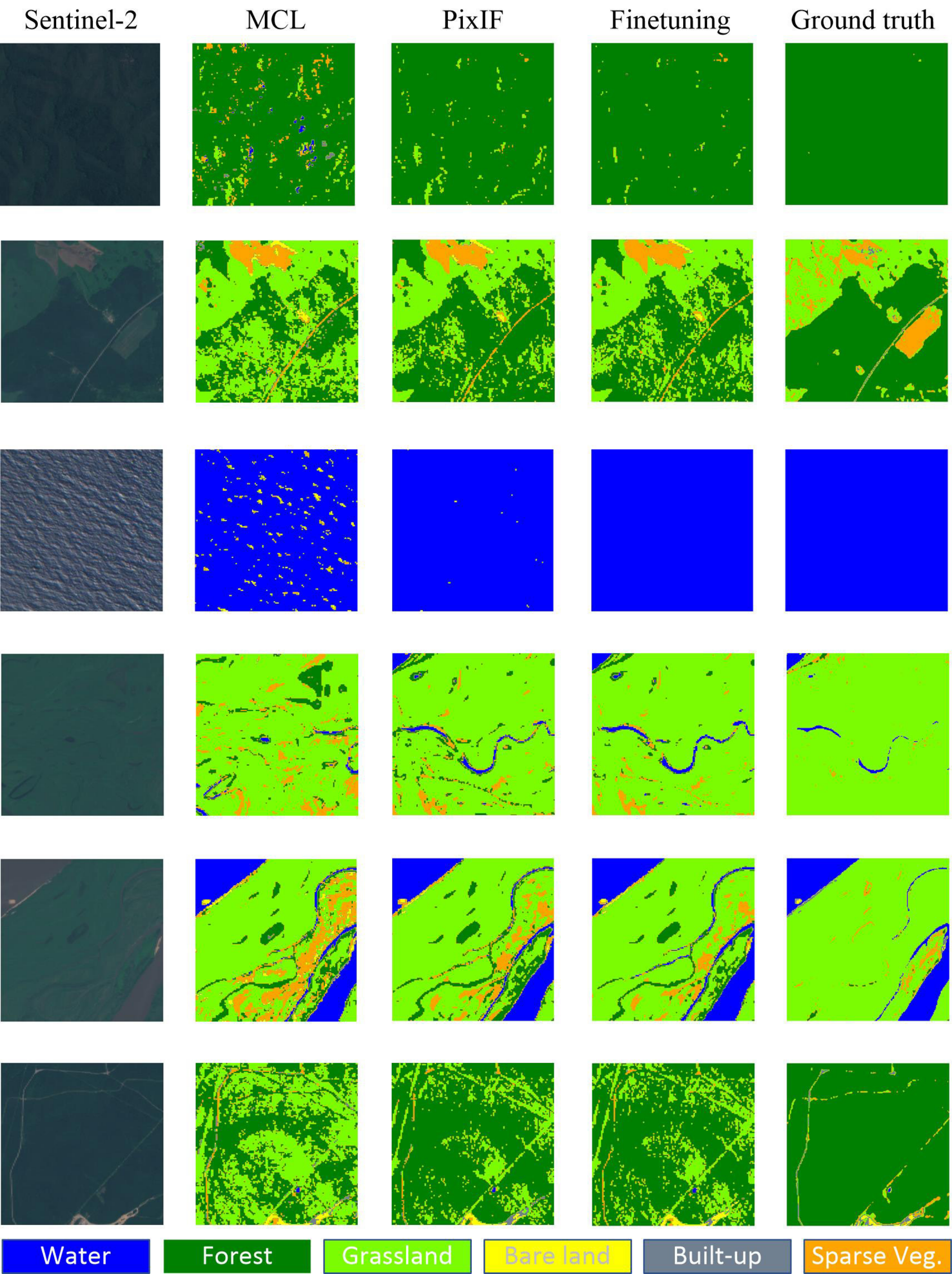}
	\caption{Self-trained land-cover maps obtained by MCL with decoder and pixel-wise feature fusion as well as its fine-tuning results.}
	\label{fig6}
\end{figure}
%o
Apart from quantitative assessment, we also made a visual quantitative comparison of the land-cover maps predicted by different methods.
%o
As shown in Fig. \ref{fig6}, the pixel-level feature fusion classifies various classes in a more accurate way, especially in small areas.
%ccc
A trend can be figured out, that is, the proposed approach achieves more smooth parsing results compared with the MCL results.  
%O
And the result of fine-tuning also leads to an improvement on AA and mIoU.
%o
Overall, the visual comparison is coherent with the quantitative results presented in Table \ref{table3} and confirms again the effectiveness of the proposed self-supervised approach on the land-cover mapping task.

\begin{table}[pt]
	\centering
	\caption{The effect of the use of geometric, photometric, shift augmentation and global loss in the proposed approach.}
	\label{table4}
	\renewcommand\tabcolsep{4.0pt}
	\centering
	\begin{tabular}{cccc|cc}
		\hline
		Only Shift & Geometric (affine) & Photometric & Global loss & AA & mIoU \\ \hline
		$\checkmark$ &   & & $\checkmark$ & 72.7 & 0.498 \\
		& $\checkmark$ & & $\checkmark$ & 68.4 & 0.460 \\
		&   & $\checkmark$ & $\checkmark$ & 68.8 & 0.464 \\
		$\checkmark$ &  & & & 70.6 & 0.479 \\ \hline
	\end{tabular}
\end{table}

\subsection{Discussion}
%O
In this section, we discuss the effects of different components of PixIF that contribute to its performance.
%O
All results showed in Table \ref{table4} are trained and tested on the same setup of the linear protocol with 1000 training samples.
%C
The first row refers to the proposed approach using only shift transformation and additional global loss.
The second row refers to the proposed approach using geometric transformation (shift, rotate, resize and sheer) instead of shift operation and additional global loss.
The third row refers to the proposed approach using photometric transformation (gaussian blur and noise) and additional global loss.
And the fourth row refers to the proposed approach only use shift operation.

%O
As one can see, the combination of shift operation and the use of global contrastive loss achieves the highest accuracy.
%O
In contrast, the lack of global loss makes the performance slightly decayed.
It demonstrates the benefits of the use of image-level contrastive loss.
%O
We also investigate a universal geometric transformation instead of shift operation in the network training, where the performance drops about 0.04 on mIoU.
%O
The proposed approach can work with the affine geometric transformation but with little performance drops.
%O
Similarly in photometric transformation, it leads to 0.03 drops on mIoU.
%O
Finally, we find that the shift operation is the most useful and simple data augmentation approach for the proposed approach in this work.

\section{Conclusion}
%o
In this paper, we proposed a new self-supervised SAR-optical data fusion approach by jointly using image-level and pixel-level contrastive loss and shift data augmentation.
%O
The proposed approach explores three fusion strategies to distill related representations from different modalities data.
%O
We additionally investigate the efficiency of SAR-optical fusion with respect to the single modality use in the proposed approach.

%o
To evaluate the performance of the proposed approach, we compared it with two instance-level self-supervised methods (i.e., CML and DCCA) and also with a weakly supervised method considering the linear protocol evaluation with different numbers of training samples.
%o
The results show that the proposed PixIF achieves the best outperforms among all self-supervised methods and achieves comparable performance to that of a weakly supervised method.
%o
The effectiveness of the proposed PixIF can be explained by the use of different levels of contrastive loss for a dense prediction task.
%o
Comparisons between the performance of proposed PixEF on SAR-optical fusion and single modality data were also considered.
%o
The experiment verified again the benefit of SAR-optical fusion in the land-cover mapping task.

%o
We also propose a self-training land-cover mapping approach based on the self-supervised pre-training and spectral information of images themselves.
%C
This approach is assessed quantitatively and qualitatively by the selected subset of DFC2020, which validated an average accuracy measure of 68\% in classifying land-cover to six classes.
%O
Results show that the proposed PixIF together with the self-training approach can achieve a comparable result on land-cover mapping without manually annotated labels.
%o
The results confirm the benefit of the proposed approach in the land-cover mapping task with respect to the use of the instance-level self-supervised methods.
%C
On the basis of the freely accessed Sentinel-1/-2 data, the proposed approach demonstrates a promising potential for automatic large-scale land-cover mapping.

%o
The proposed approach also has some limitations in initial pseudo labels generation.
%o
The considered training strategy just gives six classes and just focuses on values of backscatter values of sentinel-1 images while ignores the polarization information.
%o
Accordingly, as future development, we plan to explore the possibility of more precise classes in the presented approach for the land-cover mapping task.

\ifCLASSOPTIONcaptionsoff
  \newpage
\fi

\bibliographystyle{IEEEtran}
\bibliography{mylib}


\end{document}